\documentclass[twocolumn,journal]{IEEEtran}
\usepackage[T1]{fontenc}
\usepackage[latin9]{inputenc}
\usepackage{amsbsy}
\usepackage{graphicx}
\usepackage[unicode=true,
 bookmarks=true,bookmarksnumbered=true,bookmarksopen=true,bookmarksopenlevel=1,
 breaklinks=false,pdfborder={0 0 0},pdfborderstyle={},backref=false,colorlinks=false]
 {hyperref}
\hypersetup{pdftitle={Detection of Nonlinearly Distorted OFDM Signals via Generalized Approximate Message Passing},
 pdfauthor={Sergey V. Zhidkov},
 pdfpagelayout=OneColumn, pdfnewwindow=true, pdfstartview=XYZ, plainpages=false}
\usepackage{breakurl}

\makeatletter
 \let\oldforeign@language\foreign@language
 \DeclareRobustCommand{\foreign@language}[1]{%
   \lowercase{\oldforeign@language{#1}}}

\usepackage[caption=false,font=footnotesize]{subfig}
\usepackage{algorithmic}
\usepackage{balance}

\makeatother

\begin{document}

\title{Detection of Nonlinearly Distorted OFDM Signals via Generalized Approximate
Message Passing}

\author{Sergey V. Zhidkov,~\IEEEmembership{Member, IEEE} \thanks{S. V. Zhidkov is with Cifrasoft Ltd., Izhevsk, Russia, e-mail: \protect\href{mailto:sergey.zhidkov@cifrasoft.com}{sergey.zhidkov@cifrasoft.com}.}}

\markboth{}{Author : Detection of Nonlinearly Distorted OFDM Signals via Generalized
Approximate Message Passing}
\maketitle
\begin{abstract}
In this paper, we propose a practical receiver for multicarrier signals
subjected to a strong memoryless nonlinearity. The receiver design
is based on a generalized approximate message passing (GAMP) framework,
and this allows real-time algorithm implementation in software or
hardware with moderate complexity. We demonstrate that the proposed
receiver can provide more than a 2dB gain compared with an ideal uncoded
linear OFDM transmission at a BER range $10^{-4}\div10^{-6}$ in the
AWGN channel, when the OFDM signal is subjected to clipping nonlinearity
and the crest-factor of the clipped waveform is only 1.9dB. Simulation
results also demonstrate that the proposed receiver provides significant
performance gain in frequency-selective multipath channels.
\end{abstract}

\begin{IEEEkeywords}
OFDM, nonlinear distortion, iterative decoding, message passing, belief
propagation
\end{IEEEkeywords}

\IEEEpeerreviewmaketitle{}

\section{Introduction}

\IEEEPARstart{M}{ulticarrier} modulation in the form of orthogonal
frequency division multiplexing (OFDM) or discrete multi-tone (DMT)
has recently emerged as a preferred candidate for a wide variety of
wired and wireless communication systems. It has been used in wireless
local area networks {[}1{]}, in digital subscriber line services {[}2{]},
in optical communications {[}3{]} and is a major contender for fifth
generation mobile networks {[}4{]}. 

One of the major disadvantages of OFDM is the high crest-factor of
the time-domain OFDM waveform. To avoid in-band and out-of-band distortions,
the OFDM signal requires power amplifiers with large back-offs, which
results in reduced power efficiency. Several approaches are used to
deal with the crest-factor problem in OFDM systems. The simplest method
to reduce the crest-factor of an OFDM signal is to apply memoryless
nonlinearity, such as clipping, to the OFDM waveform before amplification
in a transmitter. This approach allows significant crest-factor reduction,
but introduces unwanted in-band distortion components. Conventional
receivers treat the distortion component as a noise and thus they
suffer from bit error rate (BER) degradation {[}5{]}. Decision-feedback
nonlinear distortion compensation schemes {[}6, 7{]} can partially
compensate for BER degradation; however, due to the error propagation
effect, they are only suitable for weak nonlinear distortions. 

In {[}8{]}, the authors theoretically demonstrated that the BER performance
of multicarrier transmission subjected to strong memoryless nonlinearity
can outperform linear transmission provided that an optimal maximum-likelihood
(ML) receiver is used. Unfortunately, an ML receiver for nonlinearly
distorted multicarrier signals with a realistic number of subcarriers
has enormous complexity. In {[}9{]}, several sub-optimal \textquotedbl{}bit-flipping\textquotedbl{}
receivers that can approach ML-performance were proposed. However,
the complexity of such receivers grows exponentially with the number
of subcarriers and modulation order, and as a result this approach
quickly becomes infeasible even for computer simulations. 

In this paper, we propose a practical receiver for multicarrier signals
subjected to a strong memoryless nonlinearity. The receiver design
is based on a generalized approximate message passing (GAMP) framework
{[}10{]}, and it allows real-time algorithm implementation in software
or hardware with moderate complexity. We demonstrate that the proposed
receiver can provide more than a 2dB gain compared with an ideal uncoded
linear OFDM transmission at BERs $10^{-4}\div10^{-6}$ in the additive
white Gaussian noise (AWGN) channel, when the OFDM signal is subjected
to clipping nonlinearity and the crest-factor of the clipped OFDM
waveform is only 1.9dB.

\section{System model }

Let us consider the system model depicted in Figure 1. In the transmitter,
${N\mathord{\left/{\vphantom{N2}}\right.\kern -\nulldelimiterspace}2}-1$
complex baseband $M$-QAM modulation symbols are transformed into
the time domain via the real inverse discrete Fourier transform (IDFT)
operation and then the cyclic prefix (CP) is added to each $N$-point
signal block. Finally, the signal is passed through a memoryless nonlinearity
block. The signal at the receiver input can be expressed as: 

\begin{equation}
y_{n}=\sum\limits _{l=0}^{L}f\left(z_{n-l}\right)h_{l}+w_{n},\quad n=0,...,N-1
\end{equation}
where $f(z)$ is a memoryless nonlinear function, $\left\{ h_{l}\right\} $
is the channel impulse response with length $L$, $w_{n}$ is the
AWGN term with zero mean and variance $\sigma_{w}^{2}$, and 

\begin{equation}
z_{n}=\Re\left\{ \sqrt{\frac{2}{N}}\sum\limits _{k=1}^{{N\mathord{\left/{\vphantom{N2}}\right.\kern -\nulldelimiterspace}2}-1}\xi_{k}\exp\left(j\frac{2\pi nk}{N}\right)\right\} ,\;n=0,...,N-1
\end{equation}
where $\xi_{n}$, $n=1,2,...,{N\mathord{\left/{\vphantom{N2}}\right.\kern -\nulldelimiterspace}2}-1$
are the complex $M$-QAM baseband symbols ($\xi_{0}=\xi_{{N\mathord{\left/{\vphantom{N2}}\right.\kern -\nulldelimiterspace}2}}=0$
to avoid DC-bias). Note that due to CP insertion $z_{-k}=z_{N-k}$,
if $k<L$. 

In the sequel, we consider the following clipping nonlinearity model: 

\begin{equation}
f\left(z_{n}\right)=\left\{ \begin{array}{l}
T,{\rm \quad if\;}z_{n}>T\\
z_{n},{\rm \quad if}\;-T\le z_{n}\le T\\
-T,\quad{\rm if\;}z_{n}<-T
\end{array}\right.
\end{equation}
where $T$ is a threshold value. Note that nonlinearity is applied
to a Nyquist-sampled signal $\{z_{n}\}$; therefore, all distortion
components fall in-band and no out-of band emission is introduced. 

In this paper, we consider the real-valued OFDM signal, sometimes
called a DMT signal, and Cartesian-type clipping nonlinearity. Nonetheless,
the results and main conclusions presented in this paper are also
applicable to complex OFDM signals and other types of nonlinearity
with the saturation effect. 

\begin{figure}[tbh]
\includegraphics[scale=0.27]{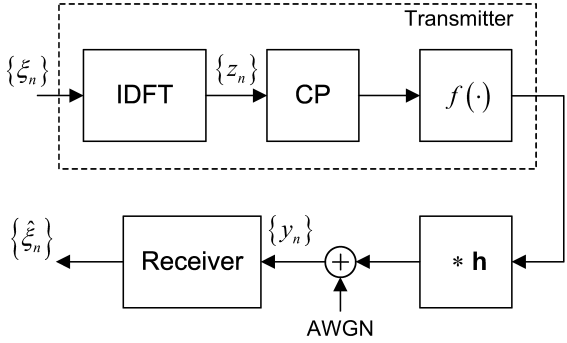}\caption{\label{fig:System-model}System model}
\end{figure}

\emph{Assumptions}: We assume that (a) the receiver knows the type
of nonlinear transfer function $f(z)$, (b) the receiver can perfectly
estimate the channel response $\{h_{k}\}$, and (c) the length of
channel impulse response is shorter than the cyclic prefix duration
($L<N_{cp}$).

\section{Receiver algorithm }

\subsection{Case I: AWGN channel }

In the AWGN channel ($h_{0}=1$, $h_{l}=0$ for $l\neq0$), the model
of the received signal can be simplified to 

\begin{equation}
y_{n}=f\left(z_{n}\right)+w_{n},\quad n=0,1,...,N-1
\end{equation}
 We can also express $\left\{ z_{n}\right\} $ as ${\bf z}={\bf Fx}$,
where $\mathbf{F}$ is a real $N\times N$ IDFT (unitary) matrix with
elements: 

\begin{equation}
F_{n,k}=\left\{ \begin{array}{l}
\sqrt{\frac{2}{N}}\cos\left({\frac{{2\pi nk}}{N}}\right),\;0\le k<{N\mathord{\left/{\vphantom{N2}}\right.\kern -\nulldelimiterspace}2}\\
-\sqrt{\frac{2}{N}}\sin\left({\frac{{2\pi nk}}{N}}\right),\;{N\mathord{\left/{\vphantom{N2}}\right.\kern -\nulldelimiterspace}2}\leq k<N
\end{array}\right.
\end{equation}
and 

\begin{equation}
x_{n}=\left\{ \begin{array}{l}
\Re\left(\xi_{n}\right),\;0\le n\le{N\mathord{\left/{\vphantom{N2}}\right.\kern -\nulldelimiterspace}2}\\
\Im\left(\xi_{n-{N\mathord{\left/{\vphantom{N2}}\right.\kern -\nulldelimiterspace}2}}\right),\;{N\mathord{\left/{\vphantom{N2}}\right.\kern -\nulldelimiterspace}2}<n<N
\end{array}\right.
\end{equation}

Model (4) is equivalent to a general problem formulation for the GAMP
algorithm {[}10{]}, which belongs to a class of Gaussian approximations
of loopy belief propagation (BP) for dense graphs. In our receiver
implementation, we use the sum-product variant of the GAMP algorithm
that approximates the minimum mean-squared error (MMSE) estimates
of $\mathbf{x}$ and $\mathbf{z}$. The GAMP algorithm adapted to
our problem is summarized below. The algorithm generates a sequence
of estimates ${\bf \hat{x}}\left(t\right)$, ${\bf \hat{z}}\left(t\right)$,
for $t=1,2,...$ through the following recursions:

\emph{Step 1) Initialization:}

$t=1$, ${\bf \hat{x}}\left(1\right)=\boldsymbol{0}$, ${\bf \boldsymbol{\mu}}^{x}\left(1\right)=\boldsymbol{1}$,
${\bf \hat{s}}\left(0\right)=\boldsymbol{0}$

\emph{Step 2) Estimation of output nodes:}

\begin{equation}
\mu_{n}^{p}\left(t\right)=\sum\limits _{k=0}^{N-1}\left|F_{n,k}\right|^{2}\mu_{k}^{x}\left(t\right),\forall n
\end{equation}

\begin{equation}
\hat{p}_{n}\left(t\right)=\sum\limits _{k=0}^{N-1}F_{n,k}\hat{x}_{k}\left(t\right)-\mu_{n}^{p}\left(t\right)\hat{s}_{n}\left(t-1\right),\forall n
\end{equation}

\begin{equation}
\hat{z}_{n}\left(t\right)=\frac{1}{C}\int\limits _{-\infty}^{\infty}ze^{-\frac{\left(y_{n}-f\left(z\right)\right)^{2}}{2\sigma_{w}^{2}}-\frac{\left(\hat{p}_{n}\left(t\right)-z\right)^{2}}{2\mu_{n}^{p}\left(t\right)}}dz,\forall n
\end{equation}

\begin{equation}
\mu_{n}^{z}\left(t\right)=\frac{1}{C}\int\limits _{-\infty}^{\infty}z^{2}e^{-\frac{\left(y_{n}-f\left(z\right)\right)^{2}}{2\sigma_{w}^{2}}-\frac{\left(\hat{p}_{n}\left(t\right)-z\right)^{2}}{2\mu_{n}^{p}\left(t\right)}}dz-\left(\hat{z}_{n}\left(t\right)\right)^{2},\forall n
\end{equation}
where 

\begin{equation}
C=\int\limits _{-\infty}^{\infty}e^{-\frac{\left(y_{n}-f\left(z\right)\right)^{2}}{2\sigma_{w}^{2}}-\frac{\left(\hat{p}_{n}\left(t\right)-z\right)^{2}}{2\mu_{n}^{p}\left(t\right)}}dz,\forall n
\end{equation}

\begin{equation}
\hat{s}_{n}\left(t\right)=\left(\hat{z}_{n}\left(t\right)-\hat{p}_{n}\left(t\right)\right)\mathord{\left/\vphantom{\left(\hat{z}_{n}\left(t\right)-\hat{p}_{n}\left(t\right)\right)\mu_{n}^{p}\left(t\right)}\right.\kern -\nulldelimiterspace}\mu_{n}^{p}\left(t\right),\forall n
\end{equation}

\begin{equation}
\mu_{n}^{s}\left(t\right)=\frac{1}{\mu_{n}^{p}\left(t\right)}\left(1-\frac{\mu_{n}^{z}\left(t\right)}{\mu_{n}^{p}\left(t\right)}\right),\forall n
\end{equation}

\emph{Step 3) Estimation of input nodes}: 

\begin{equation}
\mu_{k}^{r}\left(t\right)=\left(\sum\limits _{n=0}^{N-1}\left|F_{n,k}\right|^{2}\mu_{n}^{s}\left(t\right)\right)^{-1},\forall k
\end{equation}

\begin{equation}
\hat{r}_{k}\left(t\right)=\hat{x}_{k}\left(t\right)+\mu_{k}^{r}\left(t\right)\sum\limits _{n=0}^{N-1}F_{n,k}\hat{s}_{k}\left(t\right),\forall k
\end{equation}

\begin{equation}
\hat{x}_{k}\left(t+1\right)=\sum\limits _{m=1}^{\sqrt{M}}d_{m}P_{m,k},\forall k
\end{equation}

\begin{equation}
\mu_{k}^{x}\left(t+1\right)=\sum\limits _{m=1}^{\sqrt{M}}\left(d_{m}-\hat{x}_{k}\left(t+1\right)\right)^{2}P_{m,k},\forall k
\end{equation}
where

\begin{equation}
P_{m,k}=\frac{e^{-\frac{\left(d_{m}-\hat{r}_{k}\left(t\right)\right)^{2}}{2\mu_{k}^{r}\left(t\right)}}}{\sum\limits _{l=1}^{\sqrt{M}}e^{-\frac{\left(d_{l}-\hat{r}_{k}\left(t\right)\right)^{2}}{2\mu_{k}^{r}\left(t\right)}}},
\end{equation}
$M$ is the number of points in the signal constellation, and $\left\{ d_{m}\right\} $
is the vector of constellation points per real or imaginary component,
e.g. for 4-QAM modulation $\left\{ d_{m}\right\} =\left[-1{\rm \quad+}1\right]$
and for 16-QAM $\left\{ d_{m}\right\} =\left[-3\mathord{\left/\vphantom{-3{\sqrt{5}}}\right.\kern -\nulldelimiterspace}\sqrt{5}\quad-1\mathord{\left/\vphantom{-1{\sqrt{5}}}\right.\kern -\nulldelimiterspace}\sqrt{5}{\rm \quad+}1\mathord{\left/\vphantom{1{\sqrt{5}}}\right.\kern -\nulldelimiterspace}\sqrt{5}{\rm \quad+}{\rm 3}\mathord{\left/\vphantom{{\rm 3}{\sqrt{5}}}\right.\kern -\nulldelimiterspace}\sqrt{5}\right]$. 

The steps (7)-(17) are repeated with $t=t+1$ until $t_{max}$ iterations
have been performed. At each iteration, we calculate the Euclidean
distance between the received vector $\{y_{k}\}$, and reconstructed
time-domain waveform $f\left(\sum\limits _{k=0}^{N-1}F_{n,k}\hat{x}_{k}\left(t\right)\right)$,
and the final decision is based on vector $\left\{ \hat{x}_{k}\left(t\right)\right\} $
that corresponds to the minimum Euclidean distance $E(t)$: 

\begin{equation}
E\left(t\right)=\sum\limits _{n=0}^{N-1}\left(y_{n}-f\left(\sum\limits _{k=0}^{N-1}F_{n,k}\hat{x}_{k}\left(t\right)\right)\right)^{2}
\end{equation}

It should be noted that the integrals in (9), (10) and (11) can be
expressed in closed-form using the tabulated Gauss error functions.
However, due to the complexity of closed-form expressions, we omit
them here for clarity. In practice, (9)-(11) can be approximated using
several simple look-up tables. 

\subsection{Case II: Multipath channel }

With the addition of the multipath channel, the system model (1) deviates
from a classical GAMP estimation problem. There is an additional \textquotedbl{}channel\textquotedbl{}
sub-graph connecting nonlinear nodes $\left\{ f(z_{n})\right\} $
and observable nodes $\left\{ y_{n}\right\} $. An optimal solution
would be to incorporate the channel sub-graph into a message passing
receiver structure. However, in this paper, we resort to a sub-optimal,
yet computationally simpler approach. Namely, we introduce a channel
pre-processing step: 

\begin{equation}
{\bf y'}=IDFT\left({\bf Y}\circ{\bf H}^{\circ\left(-1\right)}\right)
\end{equation}
where ${\bf Y}$ is the DFT of $\mathbf{y}$, $\mathbf{H}$ is the
DFT of the channel impulse response $\mathbf{h}$, $\circ$ denotes
Hadamard product, and $\circ\left(-1\right)$ denotes Hadamard (element-wise)
inversion . After such a pre-processing step, we can apply GAMP algorithm
(7)-(17) to the signal $\left\{ y'_{n}\right\} $, which now can be
expressed as 

\begin{equation}
y'_{n}=f\left(z_{n}\right)+w'_{n},\quad n=0,1,...N-1
\end{equation}
where $\left\{ w'_{n}\right\} $ is the Gaussian distributed, but
non-white noise sequence. Note that we rely here on our assumption
that the channel impulse response length is shorter than the CP duration,
which allows us to model the linear convolution of $\{f(z_{n})\}$
with the channel impulse response $\{h_{k}\}$ as a cyclic convolution.
One can easily observe that (20) is essentially a well-known zero-forcing
linear equalization operation. This approach is clearly sub-optimal
since the GAMP receiver operating on $\left\{ y'_{n}\right\} $ ignores
the correlation of noise components in $\left\{ w'_{n}\right\} $.
Nonetheless, it permits low-complexity receiver implementation and
demonstrates very good performance in typical multipath scenarios,
as will be illustrated in the next section.

\subsection{Receiver complexity }

One of the advantages of the GAMP-based receiver is its moderate computational
complexity. In fact, the complexity of algorithm (7)-(17) is dominated
by the matrix transforms in (8) and (15), which can be computed efficiently
via fast Fourier transform (FFT). Thus, the complexity of the proposed
receiver is of order $O\left(4tN\log_{2}\left(N\right)\right)$, where
$t$ is the number of iterations. This compares favorably with the
complexity of maximum-likelihood receiver and sub-optimal receivers
in {[}9{]}. Overall, the receiver complexity per iteration is comparable
with that of the iterative distortion cancellers {[}6, 7{]}. 

\balance

\section{Simulation results and discussion }

The performance of the proposed decoding algorithm has been studied
by means of Monte-Carlo simulation. We simulated an uncoded OFDM system
with 2048 complex subcarriers ($N$=4096) and 4-QAM and 16-QAM modulation
with Gray mapping. The variance of $\{z_{k}\}$ was normalized to
1. For comparison, we also include here the results for a conventional
OFDM receiver and an iterative distortion canceller that attempts
to estimate and compensate for the uncorrelated nonlinear distortion
term (so called, Bussgang noise) using a decision-directed approach
{[}6, 7{]}. In all our simulations, the maximum number of iterations
($t_{max}$) was set to 30. However, it should be noted that when
the BER is less than $10^{-4}$ the average number of iterations required
for convergence is typically less than 5. The results for the AWGN
channel are illustrated in Figures 2 and 3. 

\begin{figure}[t]
\includegraphics[scale=0.6]{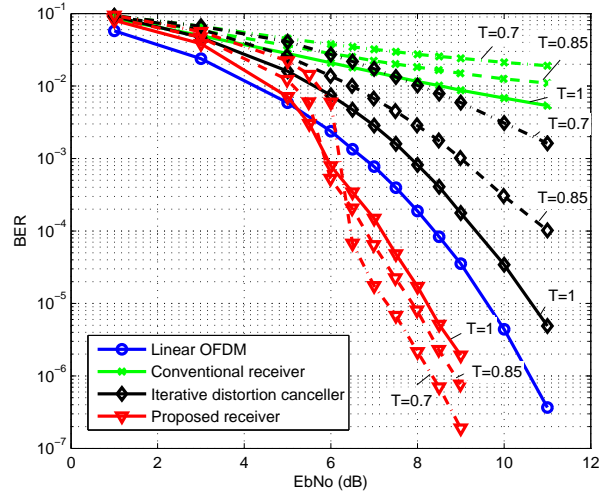}\caption{\label{fig:performance-awgn-qpsk}Performance of the proposed receiver
in AWGN channel (4-QAM, $N$=4096)}
\end{figure}

\begin{figure}[t]
\includegraphics[scale=0.6]{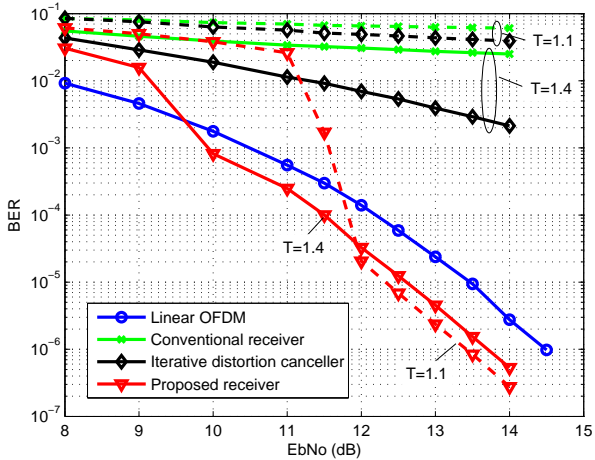}\caption{\label{fig:performance-awgn-16qam}Performance of the proposed receiver
in AWGN channel (16-QAM, $N$=4096)}
\end{figure}

As can be seen from the results presented in Figure 3, the clipped
4-QAM OFDM signal without any forward error-correction coding provides
$2\div2.3$ dB gain compared with the \emph{ideal} uncoded linear
transmission at BER=$10^{-4}\div10^{-6}$ if the clipping threshold
is set to $T=0.7$ (this value of $T$ corresponds to a clipped waveform
crest-factor around 1.9 dB). The performance gain can be as high as
2.5 dB at BER=$10^{-6}$ when $T=0.65$. At lower clipping ratios
($T\leq0.6$), the GAMP-based receiver does not converge at practically
interesting values of ${{E_{b}}\mathord{\left/{\vphantom{{E_{b}}{N_{0}}}}\right.\kern -\nulldelimiterspace}{N_{0}}}$.
At higher values of the clipping threshold ($T=0.8\div1$) the performance
improvement at low BERs is less dramatic, but the water-fall region
starts at lower values of ${{E_{b}}\mathord{\left/{\vphantom{{E_{b}}{N_{0}}}}\right.\kern -\nulldelimiterspace}{N_{0}}}$,
which results in slightly better performance at BER around $10^{-3}$. 

A less remarkable, but still significant, performance gain (>1 dB
at BER=$10^{-5}\div10^{-6}$) can also be achieved by a GAMP-based
receiver in the case of 16-QAM modulation (see Figure 3). It should
be noted, however, that in the 16-QAM case the GAMP-based receiver
does not converge at threshold values $T\leq1$. On the other hand,
lower threshold values lead to a better theoretical performance {[}8{]},
therefore, the convergence of message passing receivers for nonlinearly
distorted OFDM signals in the case of very strong nonlinear distortions
is still an open issue. 

\begin{figure}[t]
\includegraphics[scale=0.6]{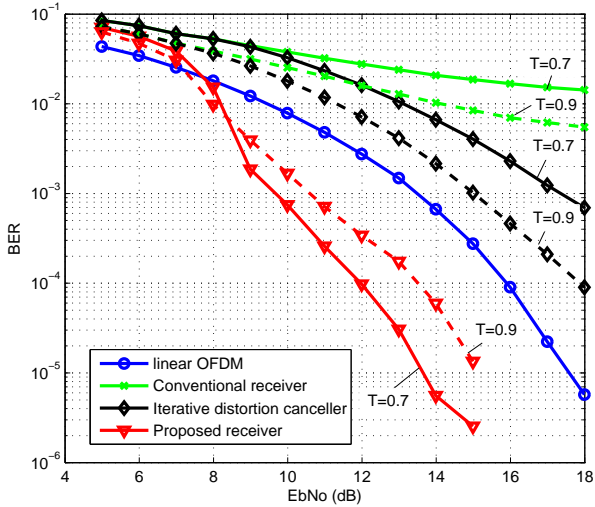}\caption{\label{fig:performance-multipath-qpsk-1}Performance of the proposed
receiver in frequency selective channel (4-QAM, $N$=4096, $T$=0.8)}
\end{figure}

Finally, in Figure 4 we illustrate the performance results for the
GAMP-based receiver with a channel pre-processing step (20) in the
frequency-selective multipath channel. The channel profile was generated
randomly from distribution $h_{k}=\mathcal{N}\left(0,1\right)\exp\left(-0.05k\right)$,
$k=0,...,63$. The frame length was set to $N=4096$, the duration
of the CP was set to $N_{cp}=64$. The results show that the proposed
receiver can achieve around $3\div4$ dB gain in such multipath channels
even with sub-optimal channel equalization processing (20).

\section{Conclusions }

In this paper, we proposed a practical GAMP-based receiver for multicarrier
signals subjected to strong memoryless nonlinearity. We demonstrated
that the proposed receiver can provide more than a 2dB gain compared
with the \emph{ideal} linear OFDM transmission at a BER range $10^{-4}\div10^{-6}$
in the AWGN channel. Note that at the same time, the crest-factor
of the Nyquist-sampled OFDM waveform is reduced from \textasciitilde{}12dB
to only 1.9dB as a result of clipping. The simulation results also
show that the proposed receiver provides significant performance gain
in frequency-selective multipath channels. Due to this fact, we believe
that, in some applications, uncoded clipped OFDM systems combined
with the proposed receiver structure can compete with coded OFDM systems.
The complexity of the proposed receiver is moderate and is suitable
for practical implementation in software or hardware. The extension
of the proposed receiver structure to coded OFDM systems and the full
utilization of the graph structure for severe frequency-selective
multipath channels is a possible venue for future research.

\end{document}